\documentclass[12pt]{article}

\usepackage{amsmath,amssymb}
\usepackage{graphicx}
\usepackage{dcolumn}
\usepackage{bm}
\usepackage[numbers,super,comma,sort&compress]{natbib}
\usepackage{color}

\begin{document}

  \makeatletter
  \renewcommand\@biblabel[1]{#1.}
  \makeatother

\bibliographystyle{apsrev}

\renewcommand{\baselinestretch}{1.5}
\normalsize


{\sffamily
\noindent {\bfseries \Large Short-time elasticity of polymer melts: Tobolsky conjecture and heterogeneous local stiffness}\\

\noindent {SEBASTIANO BERNINI${}^{1}$, DINO LEPORINI${}^{1,2}$\\}

\vspace{-8mm}

\noindent {${}^{1}$Dipartimento di Fisica ``Enrico Fermi'', 
Universit\`a di Pisa, Largo B.\@Pontecorvo 3, I-56127 Pisa, Italy}\\
\noindent {${}^{2}$IPCF-CNR, UOS Pisa, Italy\\ 
}

\vspace{-5mm}

\noindent {\em Dated: \today}\\

\begin{center}
\begin{minipage}{5.5in}

{\bf ABSTRACT:}  
An extended Molecular-Dynamics study of the short-time "glassy" elasticity exhibited by a polymer melt of linear fully-flexible chains above the glass transition is presented.  The focus is on the infinite-frequency shear modulus $G_\infty$
manifested in the picosecond time scale and the relaxed plateau $G_p$ reached at later times and terminated by  the structural relaxation. The  local stiffness of the interactions with the first neighbours of each monomer exhibits marked distribution with average value given by $G_\infty$. In particular, the neighbourhood of the end monomers of each chain are softer than the inner monomers, so that $G_\infty$ increases  with the chain length. $G_p$ is not affected by the chain length and is largely set by the non-bonding interactions, thus confirming for polymer melts the  conjecture formulated by Tobolsky for glassy polymers.


\end{minipage}
\end{center}

\clearpage

\section*{\sffamily \large INTRODUCTION}

Above the glass transition (GT) the elastic response of uncrosslinked polymer melts $G(t)$ is transient and disappears due to relaxation and viscous effects \cite{DoiEdwards}.  The decay of $G(t)$ is characterised by several regimes. In the picosecond time scale, $G(t) \simeq G_\infty$ and the elastic deformation is homogeneous and affine (${\bf r}_i \to {\bf A} {\bf r}_i + {\bf b}$ where ${\bf r}_i$, ${\bf A}$ and ${\bf b}$ are the i-th monomer position and suitable transformation matrix and vector respectively) \cite{BornHuang,ZwanzigJCP65}. In a polymer system affine motion is possible  in the limit of very small displacements only. Larger affine displacements would entail strong distortion of bond lengths and bond angles, leading to  non-homogeneous nonaffine component of the microscopic deformation to restore the force equilibrium on each monomer \cite{Theo86a,Theo86b}. Non affine motion is not specific to polymers and is also observed in crystals with multi-atom unit cell \cite{Wallace72} and atomic amorphous systems \cite{BarratPRE09,Maloney06}. 
Following the restoration of detailed mechanical equilibrium, $G(t)$ approaches the relaxed plateau $G_p$ which persist indefinitely in solids like glasses where microscopic elastic heterogeneity is revealed  \cite{BarratPRE09,DePablo04}. Above the glass transition the relaxed plateau is terminated by  the structural relaxation time $\tau_\alpha$, the average escape time from the cage of the first neighbors \cite{Puosi12,Alder87,mezardPRL10}. In polymers, for times longer than $\tau_\alpha$ the decrease of $G(t)$ is slowed down by the chain connectivity. The elastic response decays initially according to the Rouse theory, picturing each chain as moving in an effective viscous liquid \cite{DoiEdwards}. Later, the mutual entanglements between long chains  force the single chain to move nearly parallel to itself in a tubelike environment, thus ensuring additional persistence to $G(t)$ \cite{DoiEdwards,Likhtman07}.

Here, we are interested in the early "glassy" elastic regime, observed {\it above} GT at times shorter than the structural relaxation time $\tau_\alpha$.  Our interest is motivated by recent development in vibrational spectroscopy \cite{HsuVibSpectr2002} and especially Terahertz spectroscopy  which 
evidenced both  
a strikingly similar response for a wide range of disordered systems of the
dielectric response of the vibrational density of states  \cite{LunkenLoidlPRL03,ElliottJPCLett14} and coupling with mechanical properties in polymers  \cite{THz},  nanocomposites \cite{NagaiTHzApplPhysLett04,RaoPochanTHzNanocompMM07} and pharmaceuticals \cite{Zeitler2015InPress}.
We address two aspects concerning both $G_\infty$ and $G_p$ which will be compared to the features of the elastic response {\it below} GT , namely the influence of the chain-length  and  the roles played by the bonded and non-bonded interactions.

The elastic modulus of glassy polymers just below the glass transition temperature is surprisingly constant over a wide range of polymers \cite{Sperling}. In the glassy state the polymer segments largely vibrate around fixed positions on the sites of a disordered lattice and even short-range diffusion is nearly suppressed. In 1960 Tobolsky  \cite{TobolskyBook}: 
\begin{itemize}
\item noted that the elastic modulus of glassy polymers is independent of the chain length,
\item hypothesized that small strains in glassy polymers involve relative movements of {\it non-bonded} atoms, often interacting with weak van der Waals' force fields, with little or no influence by the strong covalent bonds. 
\end{itemize}

In fact, the polymers are less stiff by one or two orders of magnitude than structural metals and ceramics where deformation involves primary bond stretching \cite{HallBook89}. To a more quantitative level, Tobolsky proposed that the modulus can be evaluated to a good approximation (at 0 K) by the cohesive energy density, the energy theoretically required to move a detached polymer segment into the vapor phase \cite{TobolskyBook}. For polystyrene, a value of tensile (Young's) modulus $E = 3.3 \times 10^{9}$  Pa is calculated, which is very close to the experimental value , $3 \times 10^{9}$ Pa \cite{Sperling}. In 1974 Nielsen concluded  for unoriented polymers that the modulus in the glassy state is determined primarily by the strength of intermolecular forces and not by the strength of the covalent bonds of the polymer chain \cite{NielsenBook74}. The mechanical properties of paper offer also interesting analogies, being largely controlled by the concentration of effective hydrogen bonds and independent of both the network and the macromolecular structure, as well as  the covalent bond structure of the cellulose chain molecule \cite{CaufieldNissan06}.
Both theoretical and numerical analysis of the elasticity of glassy polymers are reported.
Yannas and Luise first separated between configtional (intramolecular) and chain-chain (intermolecular) energy barriers in a theoretical treatment of the elastic response of glassy amorphous polymers. They concluded that none of the glassy polymers studied appears to derive its stiffness predominantly from intramolecular barriers \cite{YannasLuise82}.
Linear elasticity of amorphous glassy polymers were first investigated by atomistic modelling by Theodorou and Suter \cite{Theo85,Theo86a,Theo86b}, see also ref. \cite{TheodorouMolPhys13}. It was concluded that both
entropic contributions to the elastic response to deformation and vibrational contributions of the hard degrees of freedom can be neglected in polymeric glasses, thus paving the way to estimates of the elastic constants by changes in the total potential
energy of static microscopic structures subjected to simple deformations under the requirements of detailed mechanical
equilibrium \cite{Theo86a}. More recently, Molecular-Dynamics (MD) study of deformation mechanisms of amorphous polyethylene  shows that the elastic regions were mainly dominated by interchain non- bonded interactions \cite{ElasticPEPolym10}.
The elasticity of polymer glasses has been also considered in recent MD simulations to test the predictions of the mode-coupling and replica theories of the glass transition \cite{BaschnagelEPJE11}.

The present MD study of the polymer short-time  elasticity  confirms the Tobolsky conjecture also {\it above} GT, i.e. $G_p$ is independent of the chain length and is largely set by the softer non-bonded interactions. Differently, the affine modulus $G_\infty$ increases with the chain length, mainly due to the increasing role of the stiffer bonded interactions. It is shown that the affine modulus is the average value of the local stiffness which manifests considerable distribution between the different monomers and, in particular, is weaker around the end monomers. 
It must be pointed out that: i) the MD isothermal simulations are 
carried out by varying the chain length of linear  polymers at constant density 
and not under isobaric conditions as in usual experiments and ii) the chains 
are taken as fully flexible, i.e. without taking into account
more detailed potentials accounting for, e.g., bond-bending and bond-torsions. 
Our choices facilitated the computational effort without resulting in severe 
limitations to compare the results with the experiments. Isothermal isochoric 
simulations are expected to differ from isothermal, isobaric ones only at very 
short chain length. To see this, one reminds that under isobaric conditions, 
the density increases with the chain length due to the larger fraction of the 
inner monomers with respect to the end ones, which are less well packed 
\cite{FerryBook,BarbieriEtAl2004}. 
As a rough estimate,  the additional free volume associated with a pair  of end 
monomers is about $30 \%$ of the total volume associated with two inner 
monomers (see ref. \cite{FerryBook}, page 300). This means that the number 
density $\rho(M)$ of the melt of chains with $M$ monomers is approximately 
given by $ \rho(M) \sim  \rho_{\infty} ( 1  +  2 \cdot 0.3 / M )^{-1} $ where $ 
\rho_{\infty}$ is the infinite-length density. It is seen that density changes 
due to length changes are rather small if the polymers have even few monomers. 
As to the full flexibility of the chains, one notices that the bond length of 
our model sets the length of the Kuhn segment, the length scale below which the 
chemical details leading to the segment stiffness are important 
\cite{FettersColby,Strobl,FerryBook,InoueOsakiKuhnMM96}. In practice, this 
means that each "monomer" of our model is a coarse-grained picture of the 
actual number of monomers in the Kuhn segment, namely few monomers for flexible 
or semi-flexible polymers \cite{FettersColby,InoueOsakiKuhnMM96}.

 The paper is organized as follows. In 
Sec.\ref{numerical} the MD algorithms are outlined, and the molecular model is detailed. The results are presented and discussed 
in Sec.\ref{resultsdiscussion}. In particular, Sec.\ref{gPlat} and  Sec.\ref{gInf} are devoted to the finite-frequency modulus 
$G_p$ and the infinite-frequency modulus $G_\infty$, respectively. Finally, the conclusions are summarized in Sec. \ref{conclusions}.

\section*{ \sffamily \large NUMERICAL METHODS}
\label{numerical}

A coarse-grained polymer model of a melt of $N_c$ linear fully-flexible unentangled  chains with $M$ monomers per chain is considered ( $M=3,5, 6,8,10,15,22,30,100$ ).  
The different neighbourhoods  around the inner and the end monomers of a representative chain are sketched in Fig.\ref{FigSketch}.  Non-bonded monomers at distance $r$ 
belonging to the same or different chains interact via the truncated Lennard-Jones (LJ) potential: 
\begin{equation}
U^{LJ}(r)=\varepsilon\left [ \left (\frac{\sigma^*}{r}\right)^{12 } - 2\left (\frac{\sigma^*}{r}\right)^6 \right]+U_{cut}
\end{equation}
$\sigma^*=2^{1/6}\sigma$ is the position of the potential minimum with depth $\varepsilon$, and the value of the constant
$U_{cut}$ is chosen to ensure $U^{LJ}(r)=0$ at $r \geq r_c=2.5\,\sigma$. The bonded monomers interact by a stiff potential $U^b$ which 
is the sum of the LJ potential and the FENE (finitely extended nonlinear elastic) potential \cite{sim}:
\begin{equation}
 U^{FENE}(r)=-\frac{1}{2}kR_0^2\ln\left(1-\frac{r^2}{R_0^2}\right)
\end{equation}
$k$ measures the magnitude of the interaction and $R_0$ is the maximum elongation distance. The parameters $k$ and $R_0$
have been set to $30 \, \varepsilon  / \sigma^2 $ and $ 1.5\,\sigma $ respectively \cite{GrestPRA33}. The resulting bond length
is $r_b=0.97\sigma$ within a few percent. All quantities are in reduced units: length in units of $\sigma$, temperature in units 
of $\varepsilon/k_B$ (with $k_B$ the Boltzmann constant) and time $\tau_{MD}$ in units of $\sigma \sqrt{m / \varepsilon}$ where 
$m$ is the monomer mass. We set 
$m = k_B = 1$. 

The states under consideration have monomer number density $\rho=1.086$ and  temperatures $T=0.7,1$. We  investigate the following $(N_c,M)$ pairs: (667, 3), 
(400, 5), (334, 6), (250, 8), (200, 10), (134, 15), (91, 22), (67, 30) and (20, 100), the latter for 
$T=1$ only. The pairs are chosen to ensure a number of particles $N=N_cM\approx2000$.

Periodic boundary conditions are used. $NVT$ ensemble (constant number of particles, volume and temperature) has 
been used for equilibration runs, while $NVE$ ensemble (constant number of particles, volume and energy) has been used for 
production runs for a given state point. 
The simulations were carried out using LAMMPS molecular dynamics software (http://lammps.sandia.gov) \cite{PlimptonLAMMPS}.  The model under investigation  proved useful to investigate local dynamics  \cite{AlessiEtAl01} of spectroscopic interest  \cite{Leporini94,AndreozziEtAl99,PrevostoEtAl04}.

It is interesting to map the reduced MD units to real physical units. The procedure involves
the comparison of the experiment with simulations and provide the basic length $\sigma$, temperature $\varepsilon/k_B$ and time $\tau_{MD}$͒
units \cite{sim,KremerGrestJCP90,Kroger04,PaulSmith04,SommerLuoCompPhysComm09}. For example for polyethylene and polystyrene it was found
$\sigma=5.3$ \AA, $\varepsilon/k_B = 443$ K ,$\tau_{MD} = 1.8$͒ ps and $\sigma=9.7$ \AA, $\varepsilon/k_B = 490$ K ,$\tau_{MD} =
9$ ps respectively \cite{Kroger04}. 

\begin{figure}[t]
\begin{center}
\includegraphics[width=9cm]{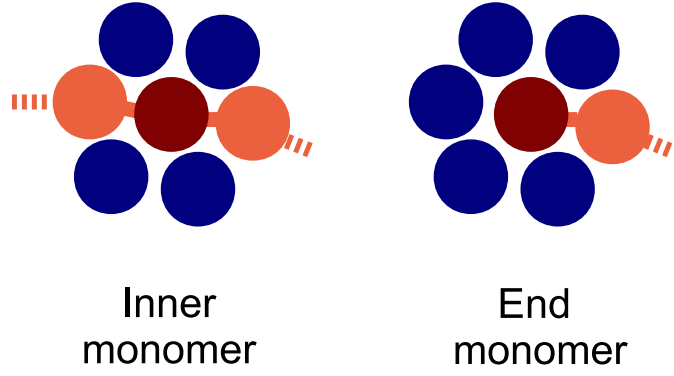}
\end{center}
\caption{Sketch of the surroundings of a tagged (red) monomer of a linear polymer. Inner monomers are bonded to other two (light red) monomers. End monomers are bonded to a single one. The different connectivity alters the arrangement of the non-bonded (blue) nearest monomers \cite{LariniCrystJPCM05,VoronoiBarcellonaJNCS14,LocalOrderJCP13,OrdineLocaleJCP15}.}
\label{FigSketch}
\end{figure}

\section*{\sffamily \large RESULTS AND DISCUSSION}
\label{resultsdiscussion}

\begin{figure}[t]
\begin{center}
\includegraphics[width=65mm]{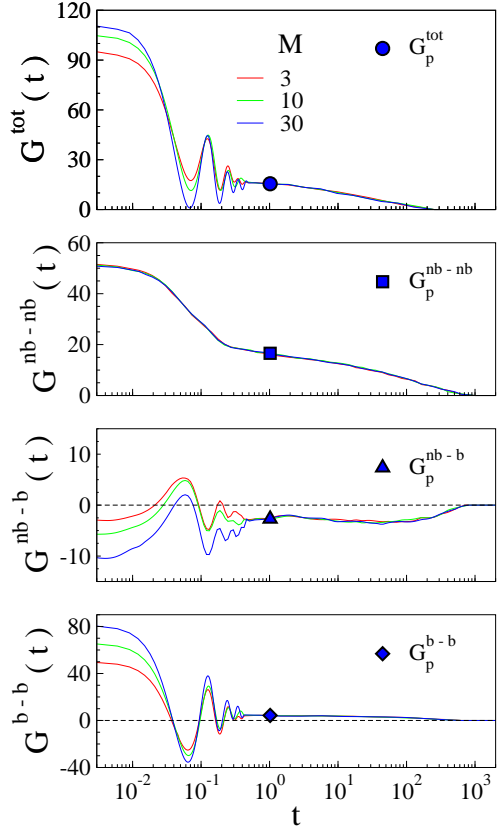}
\end{center}
\caption{Stress correlation functions for the indicated chain lengths at  $T=0.7$. The total stress correlation functions $G^{tot}(t)$, Eq.\ref{FuncPart}, is plotted in the top panel. The other panels plot the different contributions to $G^{tot}(t)$ according to Eq. \ref{FuncPart} and Eq. \ref{Glm}. The symbols mark the values of $G^{tot}_p$ and $G^{l-m}_p$ according to Eq. \ref{Gp} and Eq. \ref{GpPart}}.
\label{FigStress}
\end{figure}

\begin{figure}[t]
\begin{center}
\includegraphics[width=9cm]{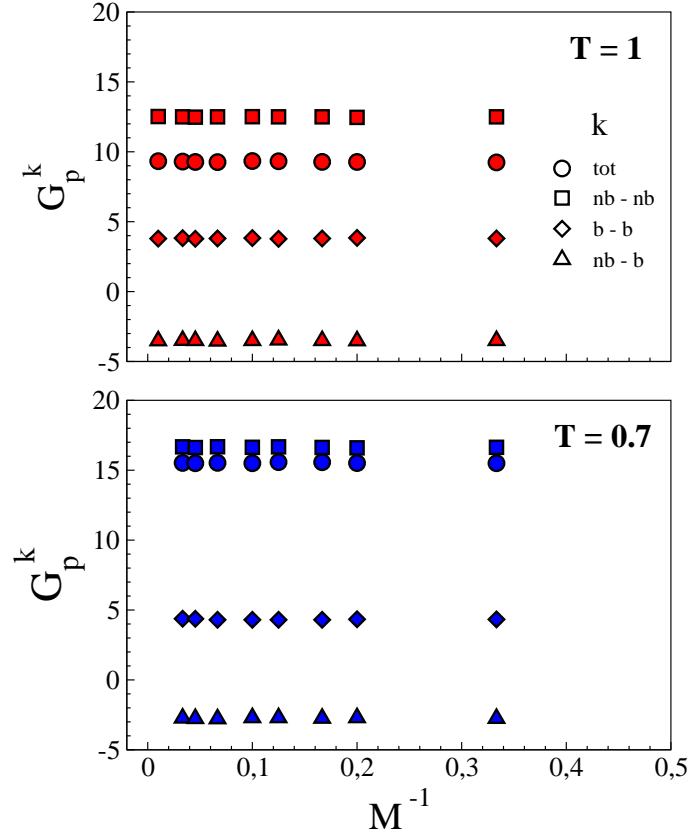}
\end{center}
\caption{Chain length dependence of the finite frequency shear modulus $G^{tot}_p$, Eq.\ref{Gp} and the related contributions $G^{l-m}_p$ with $l,m \in \{b,nb\}$, Eq.\ref{GpPart},  at the indicated  temperatures.}
\label{FigGp}
\end{figure}

\subsection*{\sffamily \normalsize Finite frequency shear modulus}
\label{gPlat}

The off-diagonal $xy$ component of the stress tensor is defined by \cite{Puosi12}:
\begin{equation}
\label{sigmaAll}
 \sigma^{tot}_{xy} = \frac{1}{V} \left[ \sum_{i=1}^N \left( mv_{x,i} v_{y,i} + \frac{1}{2} \sum_{j\neq i} r_{x,ij}F_{y,ij} 
\right) \right]
\end{equation}
where $V=N / \rho$ is the volume of the system, $v_{\alpha, i}$ is the $\alpha$ component of the 
velocity of the $i$-th monomer, $r_{\alpha, ij}$ is the $\alpha$ component of the vector joining the $i$-th monomer with the 
$j$-th one and $F_{\alpha, ij}$ is the $\alpha$ component of the force between the $i$-th monomer and the $j$-th one. 

Each monomer of the chain molecule is acted on by two distinct forces, ${\bf F}^{nb}$ and ${\bf F}^{b}$, due to the non-bonded and bonded    potentials $U^{LJ}$ and $U^b$, respectively 
(see Sec.\ref{numerical} and Fig.\ref{FigSketch} for details). In order to investigate the roles of the bonding interaction and the non-bonding LJ interaction separately, we recast $\sigma_{xy}^{tot} $ in Eq.\ref{sigmaAll} as
\begin{equation}
\label{sigmaSum}
 \sigma_{xy}^{tot} = \sigma_{xy}^b + \sigma_{xy}^{nb}
\end{equation}
with
\begin{eqnarray}
\sigma_{xy}^{b} &=& \frac{1}{V} \left( \frac{1}{2} \sum_{i=1}^N \sum_{j\neq i} r_{x,ij} F_{y,ij}^b \right) \\
\sigma_{xy}^{nb} &=& \frac{1}{V} \left[ \sum_{i=1}^N \left( mv_{x,i} v_{y,i} + \frac{1}{2} \sum_{j\neq i} r_{x,ij} F_{y,ij}^{nb}
\right) \right]
\end{eqnarray}

The shear stress correlation function is defined by \cite{ZwanzigJCP65}:
\begin{equation}
 G_{xy}^{tot}(t)=\frac{V}{k_BT} \left \langle \sigma_{xy}^{tot}(t_0) \sigma_{xy}^{tot}(t_0+t) \right \rangle
\end{equation}
where the brackets $\langle\ldots\rangle$ denote the canonical average. The average value of $G_{xy}^{tot}(t)$, $G_{yz}^{tot}(t)$ and $G_{zx}^{tot}(t)$ will be denoted as  $G^{tot}(t)$.
Note that under equilibrium \cite{ZwanzigJCP65,Boon}:
\begin{equation}
 G^{tot}_\infty  \equiv G^{tot}(0)  = G_\infty
 \label{Goo}
\end{equation}
Splitting  the total stress in bonded and non-bonded contributions as in Eq.\ref{sigmaSum} recasts the stress correlation function  as 
\begin{equation}
\label{FuncPart}
 G^{tot}(t)=\sum_{l,m \in \{b,nb\}}G^{l-m}(t)
\end{equation}
with:
\begin{eqnarray}
 G^{l-m}(t)=\frac{V}{3 k_BT}  \Big [ \left \langle \sigma_{xy}^l(t_0) \sigma_{xy}^m(t_0+t) \right \rangle + \\  \nonumber \left \langle \sigma_{yz}^l(t_0) \sigma_{yz}^m(t_0+t) \right \rangle + \left \langle \sigma_{zx}^l(t_0) \sigma_{zx}^m(t_0+t) \right \rangle  \Big ]
 \label{Glm}
\end{eqnarray}
where  $l,m \in \{b,nb\}$. 

\begin{figure}[t]
\begin{center}
\includegraphics[width=9cm]{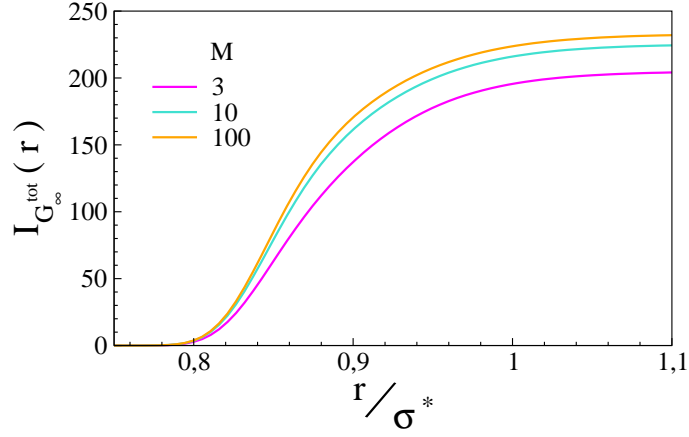}
\end{center}
\caption{Plot of the integral $I_{G_\infty^{tot}}(r) =  \int_0^r g(r') \frac{d}{dr'}\left[r'^4\frac{dU(r')}{dr'}\right] dr'$  for selected chain lengths and $T=1$. According to Eq. \ref{Ginf}, $G_\infty^{tot} =  \rho k_BT+ \frac{2\pi}{15}\rho^2  I_{G_\infty^{tot}}(\infty )$. The plot shows that $G_\infty^{tot}$ is largely due to the first neighbour shell located at $r \sim \sigma \sim \sigma^*$. For $M \ge 10$ the approximation given by  Eq. \ref{gDerivata} exceeds $G_\infty^{tot}$ by $\sim 5 \%$. }
\label{IntegGinf}
\end{figure}

Fig.\ref{FigStress}  shows the  plots the total modulus $G^{tot}(t)$ and the distinct terms $G^{l-m}(t)$ of the right hand side of Eq.\ref{FuncPart} for the states at temperature $T=0.7$ and different chain lengths. At short times ($t\lesssim0.5$) $G^{tot}(t)$ 
is characterized by oscillations with amplitude increasing with the chain length.  Inspection of the bond-bond contribution $G^{b-b}(t)$ reveals that the oscillations are due to the bond length fluctuations, affecting in part  the cross term $G^{nb-b}(t)$ too, whereas the non-bonded contribution $G^{nb-nb}(t)$ exhibits a smooth decrease at short times.
For longer times ($t\gtrsim 0.5$) the 
oscillations of $G^{tot}(t)$ vanish and both the total modulus and the distinct bonded and non-bonded contributions approach a plateau-like region. The persistence of the elastic response is due to the cage effect , namely the trapping period of each monomer in the cage of the first neighbours which is terminated by the structural relaxation time $\tau_\alpha$ (for the present states $\tau_\alpha \sim 65$ \cite{VoronoiBarcellonaJNCS14}) \cite{GotzeBook}. Beyond $\tau_\alpha$ $G^{tot}(t)$ relaxes according to the polymer viscoelasticity. We are not interested here in this long-time decay which has been addressed by other studies \cite{Likhtman07}.

To begin with, we consider the intermediate plateau region and provide a convenient definition of the plateau height. From previous work it is known that for $t\lesssim1$ the monomer explores the cage made by its first neighbors. At  $t\sim1 $ early escape events become apparent  by observing the monomer mean square displacement $\langle r^2(t)\rangle$ which exhibits 
a well-defined minimum of the logarithmic derivative quantity $\Delta (t)=\partial \langle r^2(t)\rangle / \partial \log t$ at 
$t=t^*\approx 1.02$  \cite{SpecialIssueJCP13,lepoJCP09,OurNatPhys}. $t^*$ is a measure of the monomer trapping time and is independent of the physical state in the present polymer model 
\cite{SpecialIssueJCP13,lepoJCP09,OurNatPhys}. We define the finite frequency shear modulus $G^{tot}_p$ and 
the related contributions according to Eq.\ref{FuncPart}  as:
\begin{eqnarray}
G^{tot}_p&\equiv&G^{tot}(t^*) = G_p \label{Gp} \\
 G^{l-m}_p&=&G^{l-m}(t^*), \hspace{3mm} l,m \in \{b,nb\}
 \label{GpPart}
 \end{eqnarray}
Fig.\ref{FigGp} plots the plateau height and the related distinct contributions at two distinct temperatures. It is quite apparent that: i) they do no depend on the chain lenght and ii) $G^{nb-nb}_p$  is the main contribution to $G^{tot}_p$, especially at the lowest temperature, due to the  virtual mutual cancellation of the other two contributions. Both findings fully comply with the  conjecture formulated by Tobolsky for glassy polymers \cite{TobolskyBook}. Notably, the non-bonded contribution to the plateau modulus decreases with the temperature, whereas the other contributions are nearly constant due to the stiffness of the bonds and their subsequent quasi-harmonic character.

\begin{figure}[t]
\begin{center}
\includegraphics[width=9cm]{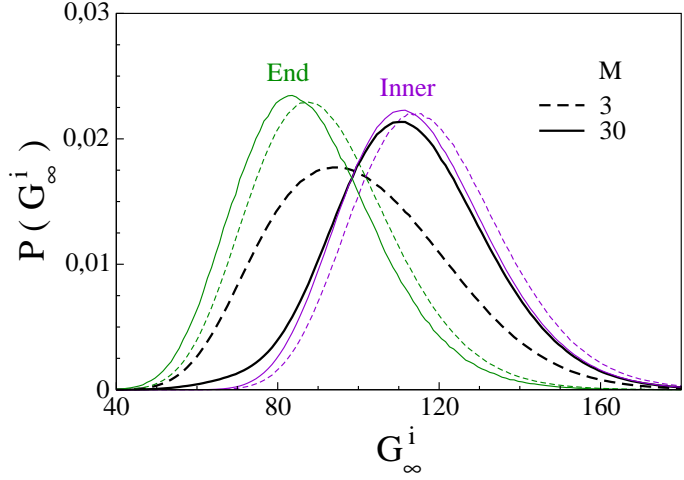}
\end{center}
\caption{Distribution of  the stiffness of the local environment surrounding 
the $i$-th monomer  $G^i_\infty$ for two different chain lengths (black lines) at $T=1$. The overall distribution  is a weighted sum of the  two components related to the end monomers (green lines), and the inner monomers (violet lines) of each chains. Note that the two components are little dependent on the chain length, and the end monomers have lower average stiffness than the inner monomers due to the lower number of bonded interactions, see Fig.\ref{FigSketch}.}
\label{DistrGinf}
\end{figure}

\subsection*{\sffamily \normalsize Infinite frequency shear modulus}
\label{gInf}

We now concentrate on the infinite-frequency shear modulus $G_\infty = G^{tot}_\infty$, Eq. \ref{Goo}, which is expressed as \cite{ZwanzigJCP65,Boon}:
\begin{eqnarray}
\label{Ginf}
 G_\infty^{tot} &=& \rho k_BT+\frac{2\pi}{15}\rho^2 \int_0^\infty g(r) \frac{d}{dr}\left[r^4\frac{dU(r)}{dr}\right] dr \\
 &\simeq& \rho k_BT+ \frac{2\pi}{15}\rho^2 \int_0^\infty r^4 g(r) \frac{d^2 U(r)}{dr^2 } dr
 \label{gDerivata}
 \end{eqnarray}
where $g(r)$ and $U(r)$ are the radial distribution function and the interaction potential, respectively.  The approximation given by Eq.\ref{gDerivata} follows by  Fig.\ref{IntegGinf} showing that the integral in Eq.\ref{Ginf} is dominated by the region of the first shell,  where $g(r)$ is maximum and the potential is close to the minimum at the investigated density and the chosen bond length.  For $M \ge 10$ Eq. \ref{gDerivata} exceeds $G_\infty^{tot}$ by $\sim 5 \%$. 

Eq.\ref{gDerivata} and Fig.\ref{IntegGinf} emphasise that $G_\infty^{tot}$ is an average local stiffness due to the interactions between  one central monomer and the closest neighbours. Thus, it is interesting to rewrite $G_\infty^{tot}$ as:
\begin{equation}
\label{Ginf1}
G_\infty^{tot} = \frac{1}{N}\sum_{i=1}^N G_\infty^i 
\end{equation}
$G_\infty^i $ has to be interpreted as a measure of the stiffness of the local environment surrounding 
the $i$-th monomer with radial distribution $g^i(r)$ :
\begin{equation}
\label{Ginf2}
G_\infty^i = \rho k_BT+\frac{2\pi}{15}\rho^2 \int_0^\infty g^i(r) \frac{d}{dr}\left[r^4\frac{dU(r)}{dr}\right] dr 
\end{equation}
Fig.\ref{DistrGinf}
plots the overall distribution of the local stiffness for two different chain lengths and compares it to  the same distribution restricted to the end  and  inner monomers. The end monomers are, on average, softer than the inner ones  due to the lower connectivity, see Fig.\ref{FigSketch}. The restricted distributions are little dependent on the chain length. Instead, the overall distribution depends on the chain length since changing the number of monomers per chain changes the relative weights of the end and the inner monomers.

\begin{figure}[t]
\begin{center}
\includegraphics[width=9cm]{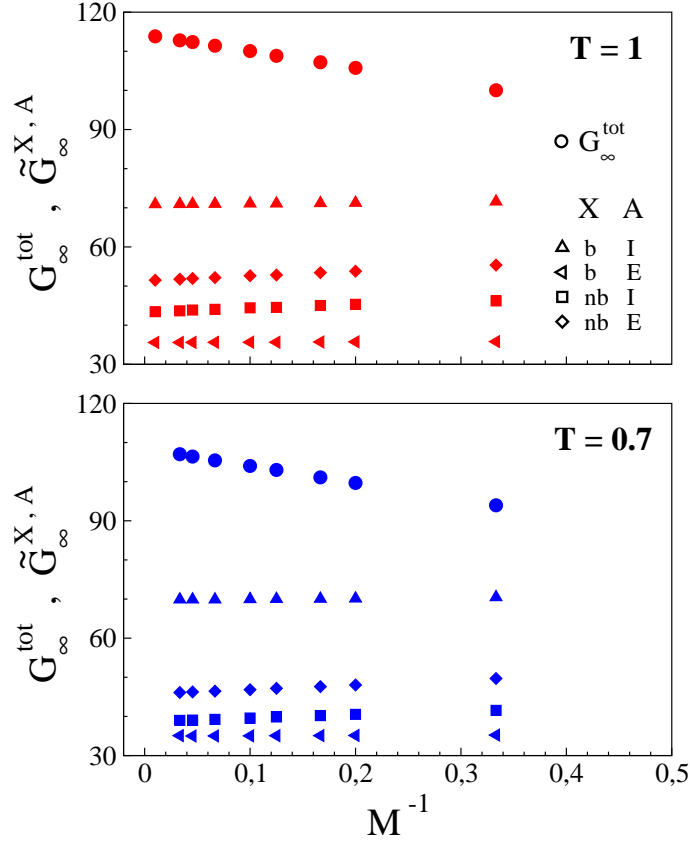}
\end{center}
\caption{Chain-length dependence of the infinite-frequency shear modulus $G^{tot}_\infty$ and the average local stiffnesses $\widetilde G_\infty^{X,A}$ with $X \in \{b,nb\},  A \in \{E,I \}$ (see Eq.\ref{Ginf6})  at the indicated  temperatures.}
\label{FigGinf}
\end{figure}

The i-th monomer is surrounded by monomers which are either bonded or non-bonded to the former with radial distributions $g^{b,i}(r)$ and $g^{nb,i}(r)$, respectively. To investigate how the bonded and non-bonded monomers affect the local stiffness we separate the two contributions:
\begin{equation}
\label{Ginf3}
G_\infty^i =  G_\infty^{b,i} +  G_\infty^{nb,i}
\end{equation}
with
\begin{eqnarray}
 G_\infty^{b,i} &=& \frac{2\pi}{15}\rho^2 \int_0^\infty g^{b,i}(r) 
\frac{d}{dr}\left[r^4\frac{dU^b(r)}{dr}\right] dr  \label{Ginf4}
\\
 G_\infty^{nb,i} &=& \rho k_BT +\frac{2\pi}{15}\rho^2 \int_0^\infty g^{nb,i}(r) 
\frac{d}{dr}\left[r^4\frac{dU^{LJ}(r)}{dr}\right] dr \label{Ginf5}
\end{eqnarray}
The average values of the bonded contributions, $G_\infty^{b,i}$, over the end monomers and the inner monomers will be denoted as  $\widetilde G_\infty^{b,E}$ and $\widetilde G_\infty^{b,I}$, respectively. The analogous averages of the non-bonded contributions, $G_\infty^{nb,i}$, will be denoted as  $\widetilde G_\infty^{nb,E}$ and $\widetilde G_\infty^{nb,I}$. In practice, the infinite-frequency shear modulus is interpreted as an weighted sum of four different kinds of average local stiffnesses:

\begin{equation}
\label{Ginf6}
G_\infty^{tot} =   \phi_I \Big [ \widetilde G_\infty^{b,I} +  \widetilde G_\infty^{nb,I}  \Big ] + \phi_E \Big [  \widetilde G_\infty^{b,E} +  \widetilde G_\infty^{nb,E} \Big ]
\end{equation}
where $\phi_I$ and $\phi_E$ are the relative weights of the inner and the end monomers, respectively:
\begin{eqnarray}
 \phi_I &=& \frac{M-2}{M} \\
 \phi_E &=& \frac{2}{M}
\end{eqnarray}
Fig.\ref{FigGinf} shows the chain-length dependence of both $G_\infty^{tot}$ and the average local stiffnesses (see Eq.\ref{Ginf6}). One notices that $G_\infty^{tot}$ increases with the chain length and the temperature, whereas $G_p^{tot}$ is independent of the chain length and decreases by increasing the temperature (see Fig.\ref{FigGp}). 
Furthermore, it is seen that $\widetilde G_\infty^{nb,I}$ and $ \widetilde G_\infty^{nb,E}$ are weakly dependent on the chain length, whereas $\widetilde G_\infty^{b,I}$ and $ \widetilde G_\infty^{b,E}$ are independent of that. In particular , it is seen that $\widetilde G_\infty^{b,I} \sim 2 \widetilde G_\infty^{b,E}$ and $\widetilde G_\infty^{nb,E} >  \widetilde G_\infty^{nb,I}$. This is due to the doubled bonded interactions of the inner monomers with respect to the end ones, and the corresponding decrease of the non bonded interactions with the first neighbours, see Fig.\ref{FigSketch}. 
The residual chain-length dependence of the non-bonded terms of $\widetilde G_\infty$ in Fig.\ref{FigGinf} is readily explained by the fact that the average density around the end monomers is lower than the one around the inner monomers \cite{BarbieriEtAl2004}. Since the monomer density is kept constant and independent of the chain length, the increase of the chain length reduces the fraction of end monomers leading to the (slight) decrease of the density around all the other monomers and the subsequent (weak) softening of the non-bonded elasticity.  Finally, we note that the temperature dependence of  $G_\infty^{tot}$ has to be ascribed to the non-bonded interactions affecting $\widetilde G_\infty^{nb,E}$ and $\widetilde G_\infty^{nb,I}$.
Fig.\ref{FigGinf} clarifies that the chain-length dependence of $G_\infty^{tot}$ is largely due to the change of the fractions of the inner and the end monomers, $\phi_I$ and $\phi_E$, rather than changes in the local stiffnesses.

\section*{\sffamily \large CONCLUSIONS}

\label{conclusions}
An extended MD study of the short-time "glassy" elasticity $G(t)$ of a polymer melt before the structural relaxation takes place has been carried out.  Two characteristic regimes are noted. In the picosecond time scale, $G(t)$ approaches the affine, infinite-frequency modulus $G_\infty$ whereas, following the restoration of detailed mechanical equilibrium, $G(t)$ approaches the relaxed plateau $G_p$ which is terminated by  the structural relaxation time $\tau_\alpha$.  

$G_\infty$ depends on the chain length whereas $G_p$ is virtually independent of that. The dependence of  $G_\infty$ on the chain length is ascribed to both the local character of  $G_\infty$, mainly set by the stiffness of the interactions with the first neighbours, and the larger connectivity, via stiff bonds,  of the inner monomers with respect to the end ones. The role of the connectivity is also exposed in the chain-length distribution of the local softness which follows by the range of different rigidity of the local environments which is fairly larger for inner monomers. 
 
$G_p$ is not affected by the chain length and is largely set by the non-bonding interactions, thus confirming also for polymer melts above the glass transition the Tobolsky conjecture originally formulated for glassy polymers.

\subsection*{\sffamily \normalsize ACKNOWLEDGMENTS}

A generous grant of computing time from IT Center, University of Pisa and Dell${}^\circledR$ Italia is gratefully acknowledged.

\clearpage



\bibliography{biblio.bib}


\end{document}